\documentclass[aps,prb,twocolumn,
superscriptaddress,
notitlepage,showpacs,floatfix]{revtex4-2} 

\usepackage{graphicx}
\usepackage{dcolumn}
\usepackage{bm}
\usepackage{xcolor}
\usepackage{amsmath}
\usepackage{amsfonts}
\usepackage{braket}

\begin{document}

\title{
Using Detector Likelihood for Benchmarking Quantum Error Correction
}

\author{Ian Hesner}
\email{ihesner@phys.ethz.ch}
\affiliation{
 IBM Quantum, IBM Research Europe -- Zurich
}
\affiliation{Institute for Theoretical Physics, ETH Zurich 8093, Switzerland}
\author{Bence Het\'enyi}
\affiliation{
 IBM Quantum, IBM Research Europe -- Zurich
}

\author{James R. Wootton}
\affiliation{
 IBM Quantum, IBM Research Europe -- Zurich
}

\date{\today}

\begin{abstract}

The behavior of real quantum hardware differs strongly from the simple error models typically used when simulating quantum error correction. Error processes are far more complex than simple depolarizing noise applied to single gates, and error rates can vary greatly between different qubits, and at different points in the circuit. Nevertheless, it would be useful to distill all this complicated behavior down to a single parameter: an effective error rate for a simple uniform error model. Here we show that this can be done by means of the average detector likelihood, which quantifies the rate at which error detection events occur. We show that this parameter is predictive of the overall code performance for two variants of the surface code: Floquet codes and the 3-CX surface code. This is then used to define an effective error rate at which simulations for a simple uniform noise model result in the same average detector likelihood, as well as a good prediction of the logical error rate.

\end{abstract}

\maketitle

\section{\label{sec:level1}Introduction}

The last few decades have seen the introduction of many quantum error correcting codes, with variants of the toric~\cite{kitaev_fault_tolerant_2003} and surface codes~\cite{dennis_topological_2002} being among the most prominent. These include the so-called rotated surface code \cite{tomita_low-distance_2014} -- now viewed as the most standard `vanilla' version -- as well as other variants that are typically tailored to certain hardware constraints or approaches to logical operators. These include those adapted to biased noise~\cite{ataides_xzzx_2021}, constrained qubit connectivity \cite{chamberland_topological_2020,kesselring_anyon_2022,mcewen_relaxing_2023,hetenyi_tailoring_2023}, implementation with Majorana qubits~\cite{hastings_dynamically_2021}, or implementation of twist defects \cite{wootton_family_2015}.

Proposals for these hardware-tailored codes are typically accompanied by a threshold plot. Numerics are performed using a simple circuit-level error model in which all components suffer depolarizing noise with probability $p$, the resulting syndromes are decoded, and the results are used to determine the threshold error rate $p_{th}$, the value of $p$ under which effective error correction is possible.

More recently, experimental implementations of the codes have also been achieved~\cite{harper2019fault,acharya_suppressing_2022, krinner_realizing_2022, sundaresan_matching_2022,
postler2022demonstration,bluvstein2023logical,gupta2024encoding,dasilva2024demonstration,hetenyi2024creating}
. Though the noise in these experiments is much more complex than that of simple error models, it is nevertheless tempting to compare results from real hardware to those from threshold plots obtained from theoretical research. In particular, to benchmark progress towards full-scale quantum error correction, we would like to know how far above or below threshold our hardware effectively operates.

One method used to do this is the $\Lambda$ factor~\cite{mcewen_leakage_2021}. This looks at how the probability of logical errors depends on the code distance and the number of syndrome measurement rounds. For a simple error model parameterized by a single noise rate $p$, this relationship depends on the factor $\Lambda^{-1} = p/p_{th}$. By looking at results from experiments at multiple code distances we can then determine the effective value of $\Lambda$ at which the system operates. Not only does this require hardware operating below threshold, but also the ability to run multiple distances of a code. A similar concept is the $\rho$ factor which similarly describes the scaling of error cluster sizes determined by a decoder, which also requires below threshold operation ~\cite{liepelt2023enhanced}.

These methods are not so effective for many cases that will be of importance in the next few years: proof-of-principle experiments with limited scope for changing system size and number of syndrome rounds, or which effectively operate above the noise threshold. It is therefore such cases that we concentrate on in this work, looking at how results from such experiments can be placed on the threshold graph in order to understand how to work towards hardware for ever more effective quantum error correction.

To demonstrate how our methods can be used in such experiments, we present results from two variants of the surface code implemented on IBM Quantum hardware. Specifically we implement the Floquet color code~\cite{hastings_dynamically_2021} and the 3CX surface code~\cite{mcewen_relaxing_2023}.

This paper is structured as follows. First we introduce the codes that will be the basis of the experimental implementations. Second, we define the average detector likelihood of a surface code as a useful intermediate parameter to connect real device data to simulation data. With this we then define the `effective $p$' of the experiment as the value of $p$ for which a simple error model yields the same average detector likelihood. Finally we look at how well the effective $p$ value can be used to predict logical error rates, and how this depends on the form of errors present in the device and the way they affect the code.

\section{\label{sec:level1}Defining the codes and their detectors}

\begin{figure}
    \centering
    \includegraphics[width = 0.49\textwidth]{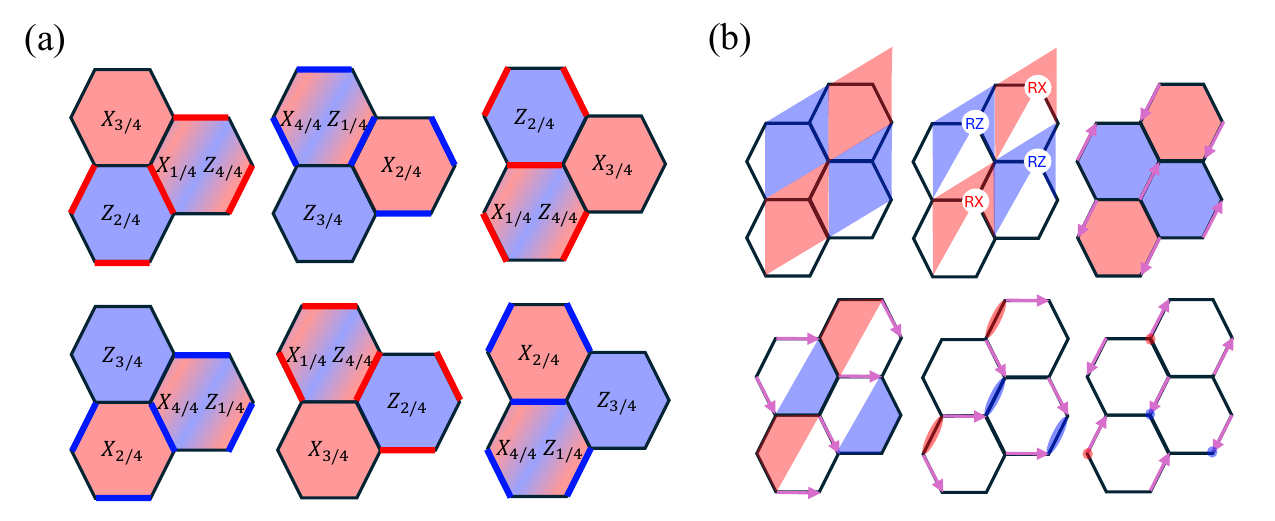}
    \caption{Time slices of the detecting regions. (a) Six sub-rounds of the Floquet code. Labels in the hexagons indicate the type of the detecting region after the X (red) or Z (blue) parity measurement with subscripts referring to the time-step relative to its creation \cite{kesselring_anyon_2022}. (b) A single round of stabilizer measurements of the 3CX code, i.e., ancilla reset followed by four CNOT operations (arrows points from control to target)  folding an X (red) and a Z (blue) detecting region onto an ancilla qubit \cite{mcewen_relaxing_2023}. Measurement of the ancilla qubits (not shown) completes the cycle. In both subfigures time progresses left to right, top to bottom and the shaded regions show the detecting regions after the operation (parity measurement or a gate).}
    \label{fig:DL_schematics}
\end{figure}

Quantum error correction is primarily implemented through the repeated application of measurements. These are chosen such that the outcomes of these measurements, or combinations thereof, have outcomes expected with certainty in the absence of noise. We will refer to these as `detectors'. Any deviation from an expected outcome is known as a `detection event', and is the result of errors. Furthermore, the locations in the circuit where an error induces a detection event is called the `detecting region' of the detector.

Let us consider the example of a quantum memory experiment of a static stabilizer code, e.g., the surface code, where the same stabilizer operators are measured in every round. Detectors in this code can be defined as the parity of subsequent stabilizer measurements with the exception of the first and last rounds of stabilizer measurements. In those rounds the definition of detectors will be determined by the preparation and the measurement of the logical state. 

For the surface code, the preparation of logical $\ket{0}$ can be performed by initializing every data qubit in the $\ket{0}$ state ensuring that the first round of Z stabilizer measurements is deterministic. In this case the Z detectors of the first round will consist of a single measurement each, while no X detectors are present. In the last round the measurement of final data qubits in the Z basis allows us to compare the last Z-stabilizer measurements to the corresponding sets of data qubit measurements. For the bulk plaquettes this involves 5 measurements per detector, one ancilla and four data qubit measurements. We note here that such temporal boundary detectors are subject to different error mechanisms compared to bulk detectors between rounds $2$ and $T-1$. Similarly, detectors on the spatial boundaries typically involve fewer qubits and therefore will report errors less frequently. Later on we define detector likelihood $\langle D \rangle$ as the probability of bulk detectors only to ensure a fair comparison of different code distances.

The two codes studied in this paper, the Floquet code and the 3CX code, have an important similarity with the surface code. Namely, every bit or phase flip in the bulk of the code induce two detection events, allowing us to use minimum-weight perfect matching for decoding. However, as opposed to static stabilizer codes, detectors are not defined by comparisons of subsequent stabilizer measurements, but rather by less intuitive sets of measurements. 

The measurement schedule for the Floquet code variant we used outlined in Ref.~\cite{kesselring_anyon_2022} and is shown in Fig.~\ref{fig:DL_schematics}(a). Defined on a hexagonal lattice, edge operators use an ancilla qubit to measure the parity of two adjacent data qubits located on the vertexes. There are three sets of edges, where each is measured in a difference basis during an overall six sub-round cycle. The measurements are such that the measured edges always commute with the existing active detecting regions defined on the plaquette faces. Each detecting region has a lifetime of 4 sub-rounds, disappearing just before an anti-commuting measurement is performed. The detecting regions exist such that every data qubit is protected by two of detectors of each Pauli type. This gives the Floquet code the same syndrome structure as a vanilla surface code, where errors create two detection events in the bulk of the code. Spatial and temporal boundary conditions are created the same way as well.

In the case of the 3CX code, four-body surface code plaquettes are measured using a modified CNOT schedule proposed by Ref.~\cite{mcewen_relaxing_2023}. The two qubits in the honeycomb unit cell are now associated with a data qubit and an ancilla respectively, while the edge qubits are unused in this protocol. The required next-nearest neighbour CNOT gates between vertex qubits of the heavy-hex were implemented as in Ref.~\cite{hetenyi2024creating}. The measurement schedule proceeds as in Fig.~\ref{fig:DL_schematics}(b). After resetting the ancilla qubit in the middle of the surface-code plaquette, the ancilla is included in the region. The CNOT gates are applied according to a specific schedule, where one of the edge directions (along the longer diagonal of the surface-code plaquette) are used twice in the schedule. Detecting regions evolve the same way as errors of the corresponding time would propagate. Once each plaquette is folded back onto a single ancilla, the ancilla can be measured in the appropriate basis. We note, that the next round of stabilizer measurements needs to be performed according to a reversed CNOT schedule because the stabilizers propagate by one lattice site in each measurement cycle~\cite{mcewen_relaxing_2023}. The second cycle is not shown explicitly in Fig.~\ref{fig:DL_schematics}(b) but one can see that ancillas initialized in the Z basis are measured on the qubit below, while X ancillas propagate down and to the left. Spatial boundary conditions provided by Ref.~\cite{mcewen_relaxing_2023} while the temporal ones are used according to Refs.~\cite{hetenyi_tailoring_2023,hetenyi2024creating}.

\section{\label{sec:level1}Detector Likelihood as an Intermediate Parameter for Error Rates}

\begin{figure}
\includegraphics[width=.48\textwidth]{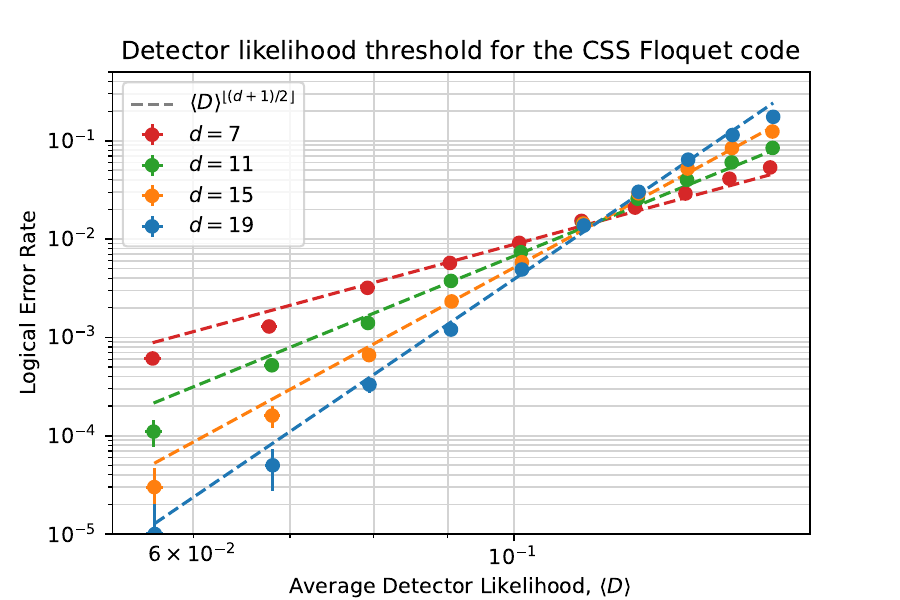}
\\
\includegraphics[width=.48\textwidth]{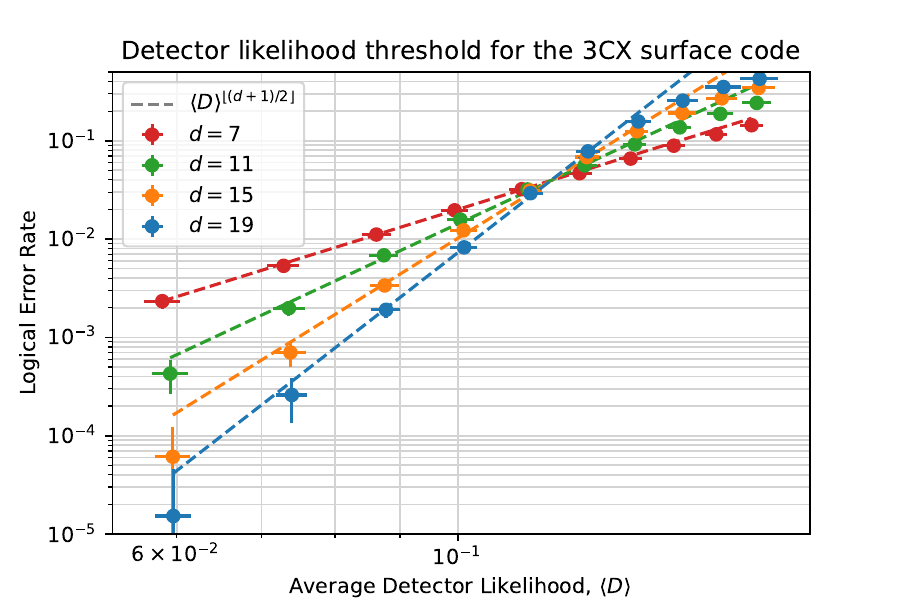}
\caption{Threshold plots as a function of average detector likelihood. The upper plot shows an average detector likelihood threshold of $\sim 11.4\%$ for the Floquet code, while the lower plot shows an average detector likelihood threshold of $\sim 11.8\%$ for a 3CX code. For the Floquet (3CX) code the simulation consisted of $2d$ subrounds ($d$ rounds). Threshold error rates are obtained by fitting the data in the subthreshold regime with the function $P_L = P_{L,0} (\langle D \rangle/D_{th})^{\lfloor (d+1)\rfloor/2}$.}
\label{III_threshold}
\end{figure}

\begin{figure}
\includegraphics[width=.48\textwidth]{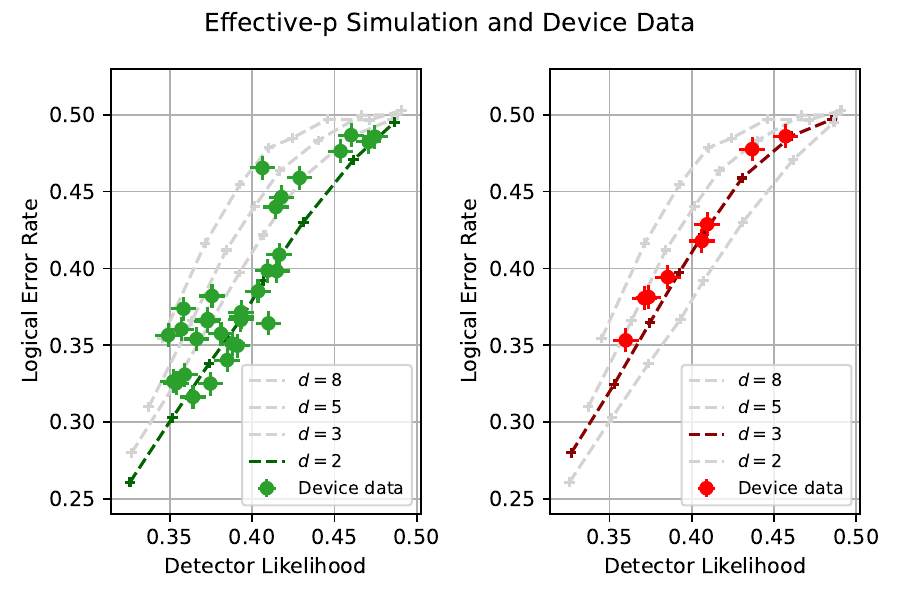}

\caption{
This figure shows how the average detector likelihood can be used to connect simulation to real device data. Simulation data is shown with darker dashed lines, while real device data is shown in lighter points with more obvious error bars. Different real device points represent different locations and dynamical decoupling schemes. Data shown here represents circuits with only 4 sub-rounds with a $|0\rangle$ initialization state. 
}
\label{III_plots}
\end{figure}

Simulating large-scale quantum processes is typically very difficult. Though efficient methods are known for quantum error correction circuits, they often have trouble simulating realistic noise. Coherent errors, cross-talk and relaxation channels are not captured by independent Pauli errors \cite{bravyi2018correcting}. For simplicity, simulations often use a single noise parameter: the probability that each element in the circuit has a failure. Alternatively, a small number of parameters may be used to reflect differences between idling, measurement and gate errors \cite{hetenyi_tailoring_2023}. In such an extended parameter space the threshold surface separates noise models that allow effective error correction from those that do not.

When individual error parameters are used in simulations, they must first be determined. This can be done indirectly, by benchmarking individual qubits and gates to measure the strength of error processes regarded as relevant for the code, which is oblivious to effects like cross-talk or the heating of the device. It can also be done using data derived from running the code~\cite{wootton_thesis_2023} using the same methods as can be used to inform the decoder of error rates~\cite{spitz_estimator_2018,wootton_repetition_2020}. In either case, the acquisition of these parameters requires assumptions on the form that the noise takes. Incorrect assumptions may result in misleading parameters, meaning that they are not necessarily an accurate measure of the noisiness of a device.

For an alternative, consider the effect that errors have on a code. They are observed in two ways: the outcomes of detectors, and the effect on logical qubits. The latter is an amalgamation of the effects of errors on all the qubits over the full duration of a quantum memory experiment. On the other hand, for LDPC codes the probability of detection events is independent of the code distance or the number of rounds. For an error model with uniform error $p$, probability of detection events will therefore be $O(p)$. These therefore provide us with a set of simple probabilities that directly measure the effects of errors on a code. These probabilities can be easily measured when running a code, either through simulations or on real hardware. As such, these probabilities can offer a useful proxy for the physical error rate that can be used to compare results from simulations and experiments. Importantly, the measured error rates incorporate the effect of cross-talk, or any other lesser known noise sources, that may depend on the details of the circuit itself.  For simplicity we will use the average over all these probabilities as our proxy. We will refer to this as the average detector likelihood, $\langle D \rangle$.

The average detector likelihood can be used in place of the physical error rate in defining a threshold plot, as shown in Fig. \ref{III_threshold}. In Figure \ref{III_threshold}(a) we see that the threshold of the Floquet code as a function of detector likelihood is between 11-12\%. Figure \ref{III_threshold}(b) shows the respective threshold plot for the 3CX code, with a detector likelihood around $12\%$. Even though the definition of detectors are quite different for the two codes, we see threshold $\langle D \rangle$ values that are very similar. Note that the threshold error rate of the Floquet and the 3CX surface codes are $\sim 0.3\%$ and $\sim 0.7\%$ respectively under the flat circuit level noise \cite{hetenyi_tailoring_2023}. We attribute this, perhaps surprising, agreement of detector-likelihood thresholds to the close ties with the surface code. Ref.~\cite{kesselring_anyon_2022} showed that the Floquet code is equivalent to a surface code in every sub-round, while the 3CX code is an even closer relative of the rotated surface code. In both cases the decoding graph is a matching-compatible graph with similar number of edges and nodes. Both codes have been decoded using \texttt{pymatching} \cite{higgott_pymatching_2021} with edge weights calculated by \texttt{stim} \cite{gidney_stim_2021}. The detector-likelihood threshold simply refers to the density of nodes the algorithm can successfully handle.

As might be expected, using such an average over all detector likelihoods does obscure some relevant details. However, as we will see, focusing on the detector likelihood for most real systems is a good enough approximation to gain valuable information about a device. In the upper plots of Fig. \ref{III_plots} we see that logical error rates for a Floquet code obtained from real device data and logical error rates obtained from our simple flat noise model simulations are associated with the same average detector likelihood. Data for 2x2 (distance 2) Floquet codes are shown on the left in green, while data for 3x3 (distance 3) Floquet codes are shown on the right in red. The darker points connected by dashed lines show simulation data with a flat noise profile of a physical error rate p. The light points that have larger error bars represent data taken from runs on a real device. Different runs represent different locations and dynamical decoupling schemes run for 4 sub-rounds. The 4 sub-round runs are shown because the lowest number of sub-rounds has the clearest signal, an effect of operating above the code's threshold. 
It can be seen that the dependence of logical error rate on function of average detector likelihood has a similar trend in both simulated and real data.

\section{\label{sec:level1}effective $p$}

\begin{figure}
\includegraphics[width=.48\textwidth]{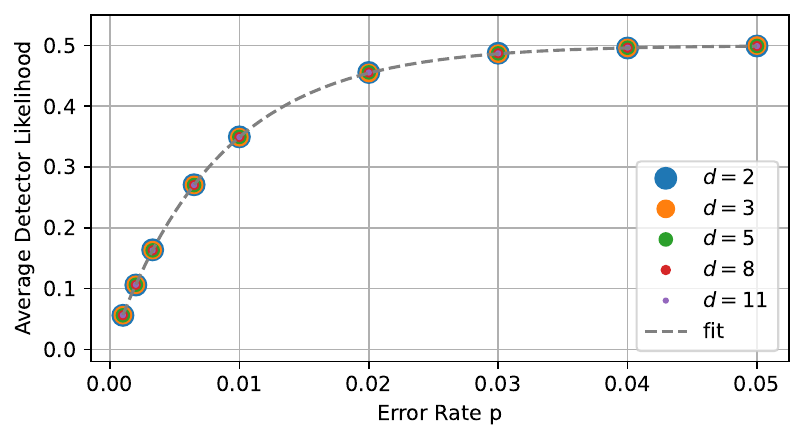}
\caption{Simulation data (shown with round markers), is fitted with Equation \ref{D_fct_p} (shown with a dashed line). This yields the appropriate $\alpha$ value in the fit to then allow us to use Equation \ref{p_fct_D} to gain an effective-p from the average detector likelihood measured on a device. Spatially and time-like truncated plaquettes are not included in the averaging.
For later calculations with Floquet codes, we took the average of these fits to get 118.7 for our alpha parameter.}
\label{IV_eff_p_fit}
\end{figure}

\begin{figure}
\includegraphics[width=.48\textwidth]{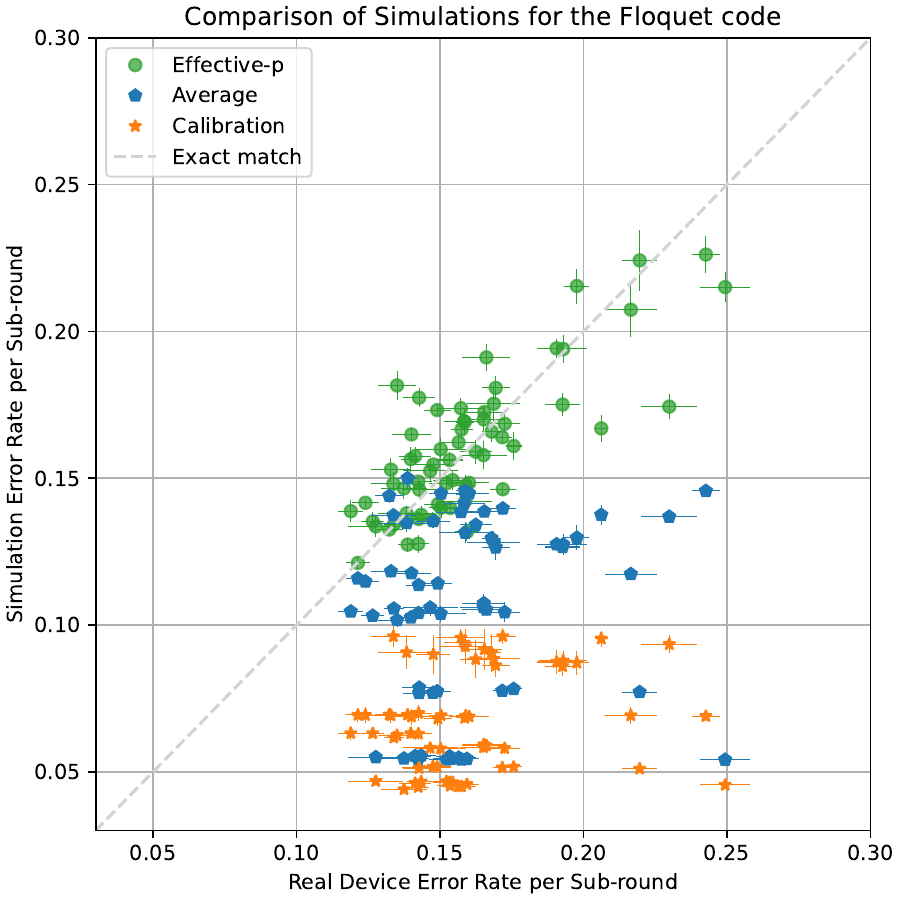}
\caption{Here data for every configuration available on \texttt{ibm\_sherbrooke} is shown for a Floquet code. On the x-axis, the real measured error rate per subround is shown, derived from the data shown in red in Fig. \ref{IV_examples}. The y-axis shows the various simulation techniques we used. Here the $p_{eff}$ simulation results are shown in blue, taking the average detector likelihood from the run with the most subrounds (16). Then the qubit specific calibration noise simulation is shown in green, while the average noise simulation is shown in orange. Parameters that are varied between points include distance, location on chip, dynamical decoupling techniques, and memory basis ($|0\rangle$ or $|+\rangle$). Clearly the $p_{eff}$ data lands much closer to exact reproduction shown with the light gray dashed line than the other techniques, implying it is more indicative of actual performance than traditional benchmarking data. 
}
\label{IV_all_sim_techniques}
\end{figure}

\begin{figure}
\includegraphics[width=.48\textwidth]{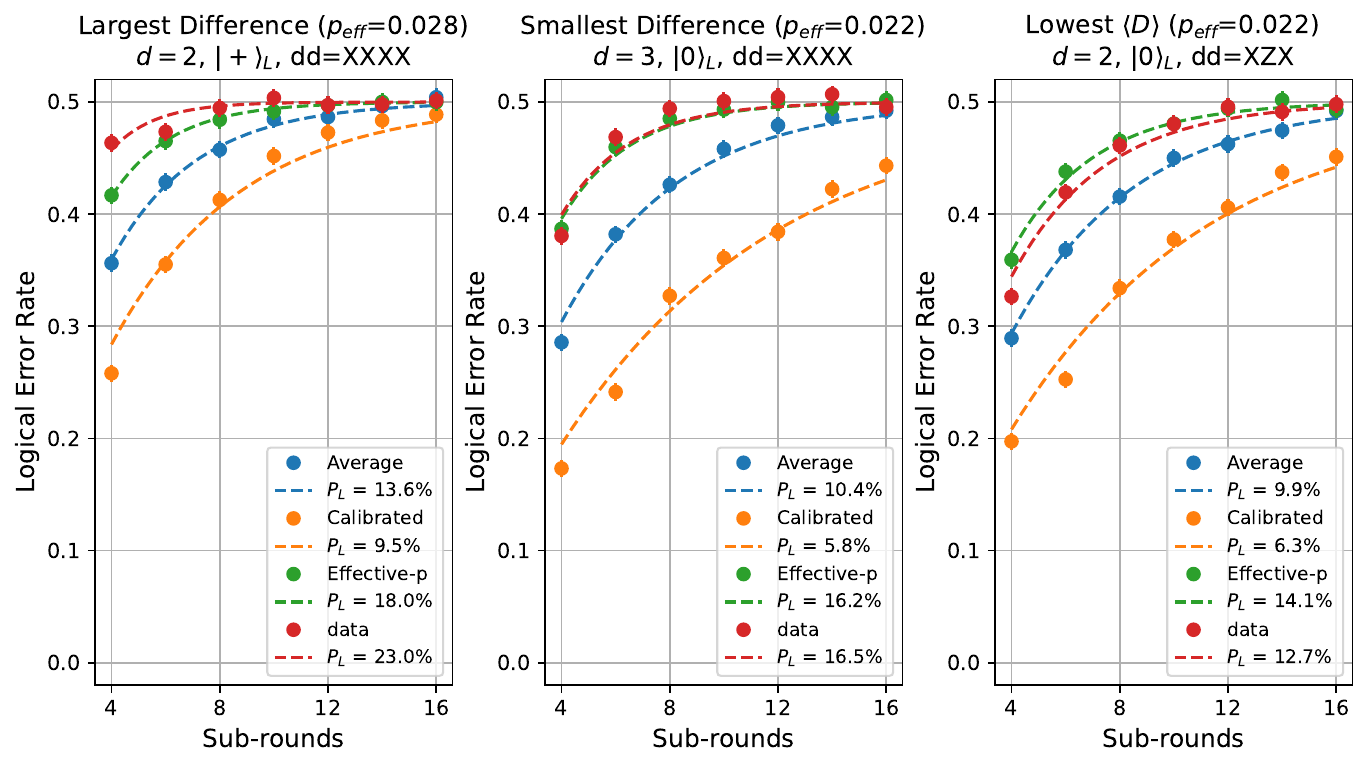}
\caption{Here we select three examples to give more detail to the data shown in Figure \ref{IV_all_sim_techniques}. The three examples include the largest and smallest difference between a $p_{eff}$ simulation and real device data, as well as the lowest overall $\langle D \rangle$. Shown on all three examples are the real device data taken in red and a $p_{eff}$ simulation in green attempting to recreate the data shown. For comparison, also shown are simulated results using calibration data from the device. Note that the results from the average error rates provide a more accurate recreation of the real device data than those using the qubit specific calibration data, offering further support for the conclusions of \cite{carroll_subsystem_2024}. 
}
\label{IV_examples}
\end{figure}

To take advantage of this correspondence, we define a new metric that we call `effective $p$' ($p_{eff}$). This is a value of $p$ for a flat noise model value which we associate as the effective noise rate of a real device. This expands on work done in references \cite{wootton_hexagonal_2022,wootton_thesis_2023} for single plaquette measurements with no encoded logical qubits. 

Given an average detector likelihood $\langle D \rangle$ extracted from a circuit run on a real device, the effective $p$ is defined simply as the $p$ value at which simulations yield the same average detector likelihood. 
Assessing all the possible fault mechanisms that can flip the detector outcome, and making crucial assumptions on the noise model (i.e., ratio of readout and gate errors), one could establish an analytical relationship between the detector likelihood $\langle D \rangle$ and some characteristic error rate $p_{eff}$. 
However, as a simpler alternative, good results can be found when fitting the data with an exponential saturation function (see App.~\ref{app:Dfit}):
\begin{equation}
    \langle D \rangle = \frac{1}{2}( 1 - e^{-\alpha p_{eff}})
    \label{D_fct_p}
\end{equation}
Here $\langle D \rangle$ refers to the average detector likelihood, and $\alpha$ is our fit parameter. This function is shown in Fig. \ref{IV_eff_p_fit}.

Once these fits have been obtained for simulated data from a given code, for any $\langle D \rangle$ value obtained from a real device running that code can be converted to an effective $p$ using the inverse of the saturation function,
\begin{equation}
   p_{eff}  = - \frac{1}{\alpha} \ln(1 - 2 \langle D \rangle).
   \label{p_fct_D}
\end{equation}

As seen previously in Fig. \ref{III_plots}, the relationship between $\langle D \rangle$ and the logical error rate for results from real device is very similar to that from simulations. As such, we can expect the logical error rate associated with $p_{eff}$ in simulations to correspond well to that in results from real devices because it recreated the same $\langle D \rangle$ as seen on the real device. A further motivation for using this $p_{eff}$ is based on the parallel between surface codes and the Ising model. The first works that introduced topological codes drew a comparison between detectors and spin flips in solid state systems \cite{dennis_topological_2002,kitaev_fault_tolerant_2003}. This comparison has been further formalized in many papers including \cite{lee_decoding_2022-1, chen_realizing_2023}. With this analogy, our approach is equivalent to supposing that just the ``spin-flip'' rate is sufficient to recreate performance of the overall system without having to worry about the details of what flipped a spin. 

This is shown to be the case in Fig. \ref{IV_all_sim_techniques} and Fig. \ref{IV_examples}. Here the predictions of the logical error rate from the effective $p$ are compared with error simulations informed by device calibration data. Figure \ref{IV_all_sim_techniques} shows the average error rate per sub-round of a Floquet code, where each point varies in device footprint, distance, logical basis, and dynamical decoupling scheme. Figure \ref{IV_examples} shows how the error rate per sub-round is obtained for three examples. We simulate the varying numbers of sub-rounds that we have real device data for, then fit the real and simulated data with Equation S3 from \cite{chen_exponential_2021} to get the logical error rate per sub-round.

For the simulations we consider the following noise sources for a given chip footprint: idle, readout, initialization, single-qubit gate, and two-qubit gate noise. All simulations were done using \texttt{stim}, and decoding was done using \texttt{pymatching}. For the ``average'' noise model, the error rate of each noise source was taken to be the corresponding error rate averaged over all the qubits of a given location. For the ``calibrated'' noise model, the individual error rates are applied to each individual qubit or connection for each noise source. For the ``effective-p'' noise model, a simple flat noise model is used. The effective-p for this simulation is derived from the $\langle D \rangle$ obtained in one of the real device runs from the real device data in Figure \ref{IV_eff_p_fit}. We only use one of the runs to show that a single benchmarking circuit is all that is needed to obtain a good effective $p$ value. We always took the run with the most numbers of sub rounds because more sub-rounds come with more detectors. Though it would be expected that simulations based on more specific benchmarking data would provide a more realistic analogue of the real noise experience by the device, we find that the logical error rates corresponding to the effective $p$ provide a much more accurate prediction. 

Note that a notion of an effective $p$ could also be defined via the logical error rate, rather than via $\langle D \rangle$. However, such an approach would arguably have several limitations. One is to limit studies to medium values of code distance. If code distance is too large and noise is above threshold, the logical error rate will quickly approach its saturation value. This will make it hard to gauge exactly how noisy the system is. If the code distance is too small, such as in \cite{wootton_measurements_2022}, it is difficult to define a logical error rate that can be reliably used. Another limitation is that the comparison becomes dependent on the details of the decoder used. Using $\langle D \rangle$, however, gives a comparison that is independent of the decoder and the definition of the logical subspace.

\begin{figure*}
\includegraphics[width = 0.49\textwidth]{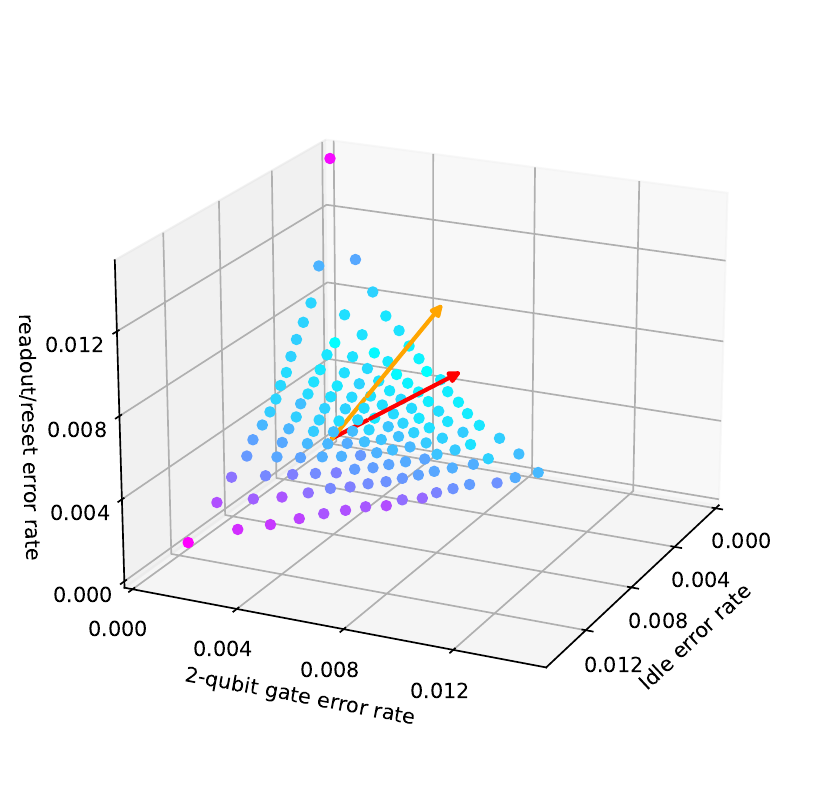}
\includegraphics[height = 0.4\textwidth]{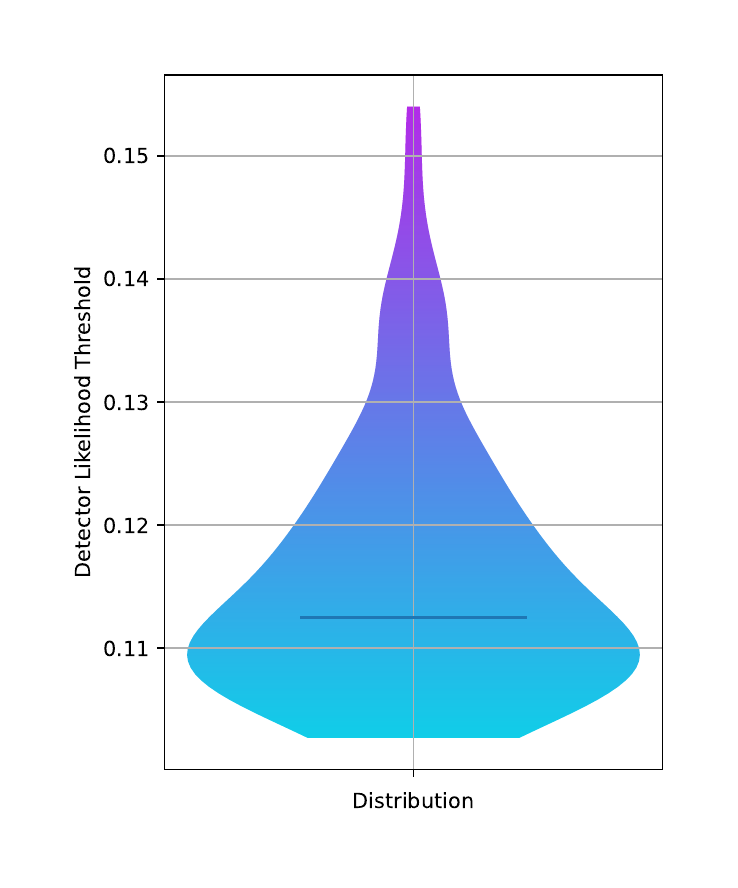}
\caption{These two figures show the variance of detector likelihood at threshold throughout a parameter space of noise parameters defined in three dimensions. To the left, a threshold surface is shown with the coloring scheme corresponding to the detector likelihood found at the threshold of the given noise breakdown. The top and the left corners can be seen as hot spots, where the detector likelihood threshold is significantly higher than in the rest of the surface. The red and orange vectors in the center of the surface represent a flat noise model and the median \texttt{ibm\_sherbrooke} noises’ locations on the surface respectively. The right violin plot shows the values of the average detector likelihood at threshold found throughout the surface. 
}
\label{III_c_surface}
\end{figure*}

The workflow for extracting an effective $p$ value is quite easy to include in existing research workflows. Modern QEC research workflows often include a step at the very beginning involving getting a running simulation. Increasingly common in the field, and also done here, this is done by leveraging \texttt{stim} and \texttt{pymatching} and achieving a threshold graph. This is a beneficial check for two reasons. First, having a functional \texttt{stim} circuit means that the commutation relations between all detectors and/or logical operators are functioning correctly. If that were not the case, it would be evident from error messages provided by \texttt{stim}. Second, plotting a threshold graph verifies that the circuit being explored is implemented in a fault-tolerant fashion. Because this simulation infrastructure is frequently already in place, defining an effective $p$ is quite easy from there. First simulations done with a flat noise model can be used to create a plot like in Figure \ref{IV_eff_p_fit} comparing the average detector likelihood to the flat physical noise level $p$. For our approach, we only include full-sized detectors, removing the space-like or time-like truncated plaquettes from the average. Not removing truncated plaquettes makes the fit in Figure \ref{IV_eff_p_fit} depend on number of subrounds for including time-like truncated plaquettes, or the size of the code for including space-like truncated plaquettes. The simulations run then are fitted with Eq.~\eqref{D_fct_p} to gain the fitting parameter $\alpha$. This $\alpha$ can then be used with Eq.~\eqref{p_fct_D} to get an effective $p$ value from the average detector likelihood gained from a real device.

\section{\label{sec:level1}Limitations of $p_{eff}$}

\begin{figure}
\includegraphics[width=0.5\textwidth]{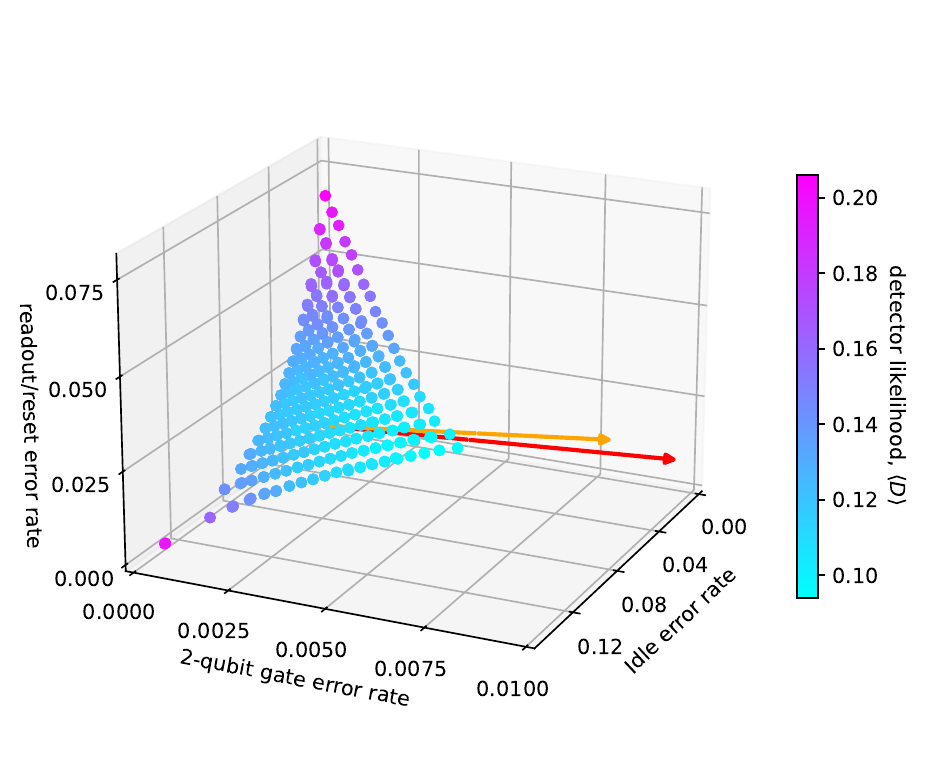}
\caption{Similar to Figure \ref{III_c_surface}, this figure shows the threshold surface for a 3CX code. Notice it is very slanted in this parameter space of noise parameters, particularly so because making a 3CX code fit on a Heavy-Hex lattice requires many additional CNOT gates. It is for this reason that comparison of Fig. \ref{III_threshold} is done against a simulation with a biased noise model, rather than a flat noise model.}
\label{V_VSC_surface}
\end{figure}

It is important to note that the accuracy of this approximation is subject to the noise breakdown of a device which we define as a set of parameters, gate-, readout- and idling error rates, used in the simulations, e.g., the flat noise model has a noise breakdown $(p,p,p)$ for a given error rate $p$. Though we have seen that it is very successful when benchmarking one of the IBM Quantum devices, in the cases of extreme noise bias the accuracy of the approximation will be reduced. E.g., in a noise model with only readout errors on the ancillas, no logical errors could happen, but the detector likelihood could be arbitrarily high. For a more realistic and quantitative picture we analyze the error threshold and the detector likelihood at threshold for different noise biases.

Figure \ref{III_c_surface} shows a threshold surface of the Floquet code on the left hand side. A threshold surface is created by calculating the threshold for different directions in a 3D parameter space of noise parameters, i.e., different noise breakdowns, then plotting the threshold as the radial magnitude. The specific shape of these surfaces depends on the error correcting code, as shown in \cite{hetenyi_tailoring_2023}. The location of the flat noise model in this parameter space of noise parameters is marked with a red vector, while average \texttt{ibm\_sherbrooke} noise breakdown is shown in orange. The coloring corresponds to the detector likelihood found at the threshold of the given noise breakdown. The top and the left corners can be seen as hot spots, where the detector likelihood threshold is significantly higher than in the rest of the surface. This means that the slopes and positions of the simulation data will be shifted by this change in threshold. Here we can see that the the detector likelihood threshold for \texttt{ibm\_sherbrooke} is quite similar to that of the flat noise model. This is because the Floquet code threshold surface is quite flat between the location of the orange vector, representing our device, and the red vector, representing where a perfect $p_{eff}$ would be. 

\begin{figure}
    \centering
    \includegraphics[width = 0.48\textwidth]{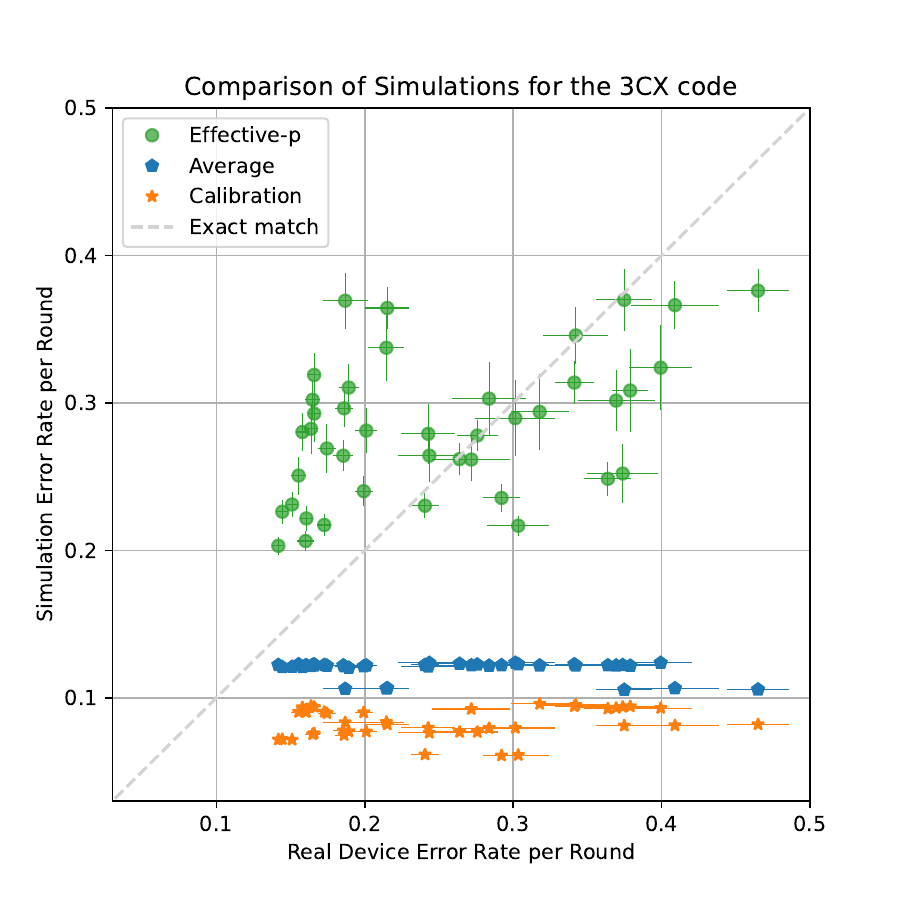}
    \includegraphics[width = 0.48\textwidth]{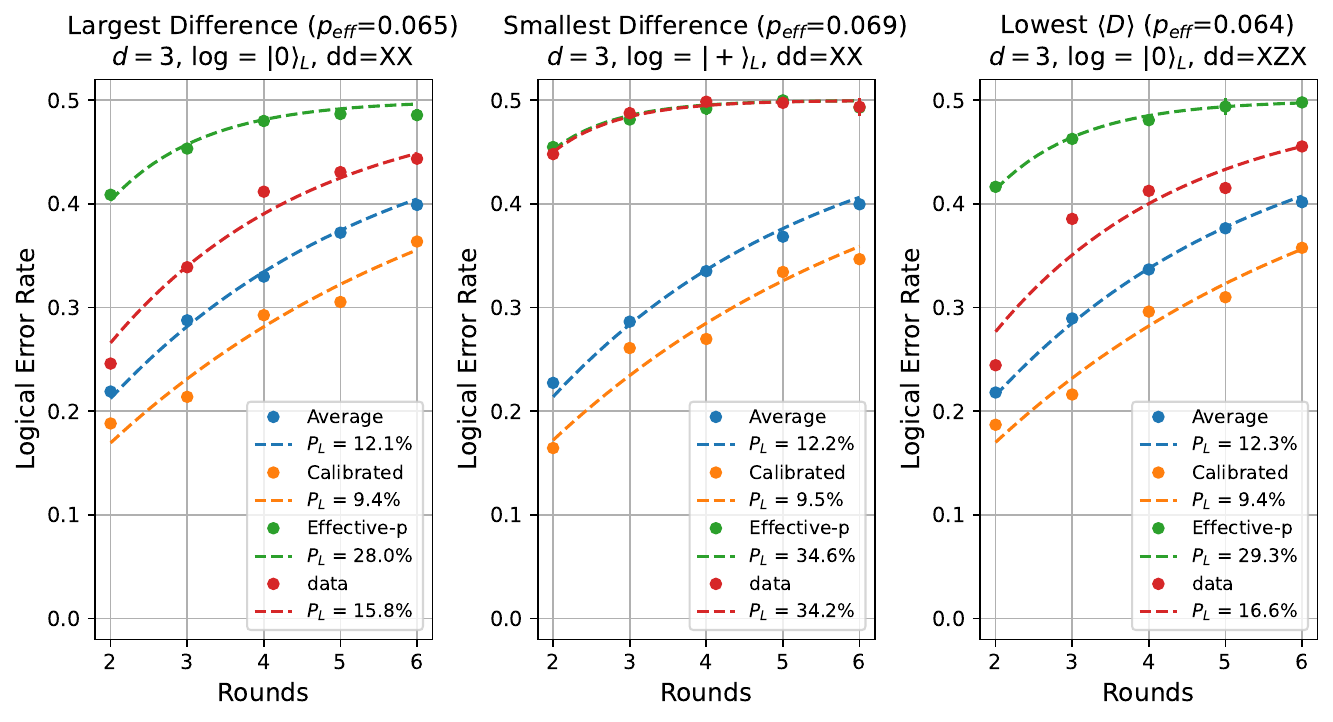}
    \caption{3CX data taken on sherbrooke. Error rates used for weighting the decoding graph and for the noise breakdown for the logical error vs detector likelihood are taken roughly from the mean calibration data}
    \label{3cx_per_round}
\end{figure}

With different codes, this threshold surface can be sloped or contain non-linearities. As an example of such an anisotropic threshold surface, Fig. \ref{V_VSC_surface} shows the that of the 3CX code. Although the detector likelihood threshold changes significantly between different noise breakdowns, the detector likelihood remains a useful figure of merit for optimizing quantum devices since the function is slowly changing in the vicinity of a given direction. E.g., in Fig. \ref{V_VSC_surface}, the noise breakdown of \texttt{ibm\_sherbrooke} corresponds to a similar detector-likelihood threshold as the flat noise model.
Furthermore, the analysis of the 3CX data on Fig. \ref{3cx_per_round} still shows an improved characterization when using effective $p$ compared to average error rates or calibration data.

\section{\label{sec:level1}Conclusion}

Here we outline a new benchmarking strategy for the performance of quantum hardware when implementing quantum error correction. This is done by measuring the average detector likelihood of the device when implementing a code. This is a metric which implicitly takes into account noise sources not captured by the standard circuit level noise model, and even noise sources missing from more complicated noise models, because every possible physical noise channel on a real device contributes to the triggering of a detector. From this we define an effective $p$, describing the equivalent error rate for a simple simulated noise model.

Our methods are applied to experimental implementations of the Floquet and 3CX codes. These show that the approach can provide excellent predictions of the logical error rates when the noise is not too biased, but will be increasingly approximate for extreme noise bias or codes with unfriendly threshold surface topologies.

\section{Data Availability}
The python scripts used for the analysis, generating the \texttt{Qiskit} and \texttt{stim} circuits used for the data collection and the simulations, as well as the data itself will be made publicly available.

\section{Acknowledgements}  \label{sec:acknowledgements}

The authors are grateful for the donation of IBM Quantum resources to complete this project. The authors also thank Malcolm Carroll for critical reading of the manuscript.

BH acknowledges support from the NCCR SPIN, a National Centre of Competence in Research, funded by the Swiss National Science Foundation (grant number 51NF40-180604).

JRW was sponsored by the Army Research Office and was accomplished under Grant Number W911NF-21-1-0002. The views and conclusions contained in this document are those of the authors and should not be interpreted as representing the official policies, either expressed or implied, of the Army Research Office or the U.S. Government. The U.S. Government is authorized to reproduce and distribute reprints for Government purposes notwithstanding any copyright notation herein.

\newpage

\appendix

\section{\label{sec:level2}The Floquet Code as a Surface Code}

Floquet codes were originally developed by Hastings and Haah in Ref.~\cite{hastings_dynamically_2021} for a honeycomb lattice with periodic boundary conditions. It can be seen as a subsystem code version of the Kitaev's honeycomb lattice model~\cite{kitaev_anyons_2006,poulin_stabilizer_2005}, for which the logical operations evolve throughout the course of the syndrome measurement rounds.

The non-trivial time dynamics of the logical operators is a key feature of the Floquet codes. The logical operators move around the code and can even change Pauli type through their evolution. The specifics of the movement and Pauli type of the logical operator, as well as the lifetimes and Pauli type of the code's detectors, are determined by the specific measurement scheduling. As long as the appropriate commutation relations are fulfilled, Floquet codes are very flexible when it comes to measurement scheduling. There have been a lot of variations of measurement scheduling and border constructions for planar implementations, including \cite{gidney_benchmarking_2022, paetznick_performance_2023, kesselring_anyon_2022}. Here we use the CSS-style defined in \cite{kesselring_anyon_2022}. 

Though the dynamic aspects of Floquet codes occlude some of the underlying structure, it can be shown that they are simply a variant of the surface code. This can be most easily seen by considering an anyonic interpretation of a syndrome \cite{brown_poking_2017}. The Pauli frames for both anyon types (e and m) are defined by those of the respective logical operators. Then, by applying that Pauli frame to the detectors of the code, it can be seen that all detectors act as either e- or m-type detectors. It turns out that every data qubit in the bulk is protected by a pair of each anyonic detector type, just like in the typical vanilla surface code. This results in all of the properties of the code acting the same as any other surface code when viewed from an anyonic perspective, including boundary construction and decoding strategies \cite{haah_boundaries_2022}.

To our knowledge, the only previous implementation of a Floquet style circuit on a real device was done in \cite{wootton_hexagonal_2022,wootton_measurements_2022}. These ran Floquet-style syndrome measurements, but without the boundary conditions required to support a logical qubit.

\section{\label{sec:level2}3CX Code}

Similarly to the Floquet codes, the 3-CX surface code cannot be described as a static stabilizer code. In order to get a better understanding of the time dynamics of stabilizer measurements, a new tool referred to as the `detecting region formalism' was developed in Ref.~\cite{mcewen_relaxing_2023}, similar to Pauli webs developed in \cite{bombin_unifying_2023}. In this picture stabilizers can also have support on ancillas (e.g., a one-qubit Z stabilizer emerges after the initialization of the ancilla), and the stabilizer generators propagate through gates and measurements according to their respective stabilizer flows \cite{gottesman_opportunities_2022}.

Such a picture opens several opportunities to modify the syndrome extraction cycle of the surface code without sacrificing the low connectivity and high threshold achievable with minimum-weight perfect-matching. In the 3-CX variant proposed by Ref.~\cite{mcewen_relaxing_2023}, only three of the four two-qubit connections are used (one of them twice) in each cycle. The schedule of CX gates are chosen in a way that the code can be implemented on a device with hexagonal rather than a square connectivity graph.

The peculiar CX schedule results in the original X (Z) stabilizers migrating from up to down (left to right). Consequently, a bulk detector measurement outcome of a given ancilla needs to be compared to an outcome of a neighbouring ancilla measurement at a later time. To accommodate these dynamics and respect the reduced connectivity, the spatial boundary conditions of the original surface code need to be adjusted such that X (Z) type ancillas are only present on the lower (right) boundary \cite{mcewen_relaxing_2023}. Importantly, these adjustments of the original surface code do preserve syndrome structure of the code.

The dynamics of the stabilizers, however, affect which detectors should be included in the detector likelihood estimate. For example, stabilizers that emerge at the boundary and migrate to a bulk plaquette cover a smaller space-time volume and therefore have less opportunities to detect errors. A detector likelihood estimate that is independent of the code distance can therefore only include detectors where bulk plaquette migrates to another bulk plaquette.

\section{\label{sec:level1}Point-like noise}
\label{appendix:point_like_noise}

\begin{figure}
    \includegraphics[width=.3\textwidth]{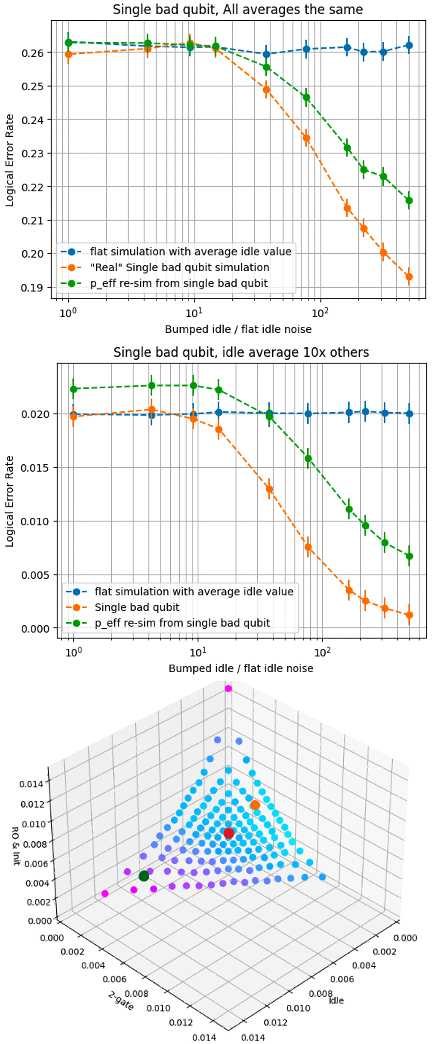}
    \caption{The above plots show simulations for varying levels of a point like noise profile. The top plot has an overall flat noise breakdown (represented by a red dot on the threshold surface below), while the second plot shows a noise bias of average idle noise / other noise = 10, represented by a dark green dot.}
    \label{A_point_noise}
\end{figure}

In addition to exploring the effectiveness of $p_{eff}$ throughout the average parameter space of noise parameters defined in the threshold surface, we also did a brief exploration of spatially variant noise within a specific noise channel. Specifically, we looked into what happens if a single data qubit in the center of the code has a much worse idle error rate than the rest of the qubits. To explore this we set all qubits to the same flat noise value, then bumped the idle noise of a single data qubit close to the center of the code. To make sure that the simulation does not move around the noise surface, which would affect the effective $p$ fitting as seen in the main text, we increased this flat noise while decreasing the bumped noise to maintain the same overall average idle noise for the simulation. Starting values were $10^{-3}$ for the flat noise and 0.5 for the bumped noise. Then, each data point brings the two noise values closer together while maintaining the same overall average. Other noise channels were all set to the same value, either the average idle noise or the average idle noise divided by ten to do a comparison of different locations in the noise breakdown space. Both of the results can be seen in Figure \ref{A_point_noise}. When all averages are the same, i.e. in the center of our threshold surface at the red dot, then effective $p$ should be working optimally from a noise breakdown perspective. As can be seen, when there is extreme point-like noise, effective $p$ does a much better job characterizing our `real device' than using average noise models. Having a large noise bias toward the idle noise moves the noise breakdown to the green dot seen on the noise threshold. This is where effective $p$ techniques should be at its worse. While the overall dynamics of changing the noise landscape does a good job, the effective $p$ is shifted up in logical error rate by quite a bit. At relatively flat noise landscapes, where bump / flat noise is low, the average noise model approach works better, but effective $p$ still gives a better estimation at high point-like noise because it still traces the shifting dynamics of a spiky noise landscape better than the averaging approach.

\section{Detector likelihood fitting formula}
\label{app:Dfit}

Taking a set of measurements as a detector, one may count every possible independent noise mechanisms that would change the parity of the measurement results. For a total of $N$ such error mechanisms we define $p_i$, the probability that the $i$th mechanism triggers the detector. (Note that e.g., depolarizing error triggers a detector with $2p/3$ probability). The probability that a detector detects an error event is 
\begin{equation}
    \langle D \rangle = \sum_{k \in \text{odd}}^N 
    \sum_{
        \substack{
            \{n_i\}|\\
            \sum n_i = k
        }
    }
    \prod_{i=0}^N p_i^{n_i} (1-p_i)^{1-n_i}.
    \label{appeq:Dexact}
\end{equation}
From this formula, it is clear that $\langle D \rangle$ in general is a polynomial of order $N$ ($N-1$) for odd (even) $N$.

Assuming $p_i=p$ we can rewrite the formula above as
\begin{equation}
    \langle D \rangle = \sum_{k \in \text{odd}}^N {N \choose k} p^{k} (1-p)^{1-k}.
\end{equation}
This can be easily traced back to a well-known result in error correction. Namely
\begin{equation}
    1-2\langle D \rangle = (1-2p)^N,
\end{equation}
from which we get
\begin{equation}
    \langle D \rangle = \frac{1-(1-2p)^N}{2} \approx \frac 1 2 (1-e^{-2Np}).
\end{equation}
While strictly speaking $p_i = p$ is not justified, the monotonity of the polynomial in Eq.~\eqref{appeq:Dexact} ensures that the exponential approximation remains close to the exact formula.

\bibliography{bib_1026}

\end{document}